\documentclass[aps,prd,twocolumn,nofootinbib,showpacs,superscriptaddress]{revtex4-1}

\usepackage{amssymb}
\usepackage{graphicx}
\usepackage{amsmath, amsthm}
\usepackage{epstopdf}
\usepackage{hyperref}

\usepackage{color}

\newcommand{\be}{\begin{equation}}
\newcommand{\ee}{\end{equation}}

\begin{document}

\title{Hawking-Hayward quasi-local energy under conformal 
transformations}

\author{Angus Prain}
\email{aprain@ubishops.ca}
\affiliation{Physics Department and STAR Research Cluster, 
Bishop's University,
Sherbrooke, Qu\'ebec, Canada J1M~1Z7
}

\author{Vincenzo Vitagliano}
\email{vincenzo.vitagliano@ist.utl.pt}
\affiliation{  CENTRA, Instituto Superior T\'ecnico,  
Universidade de Lisboa, 
Avenida Rovisco Pais 1, 1049 Lisboa, Portugal
}

\author{Valerio Faraoni}
\email{vfaraoni@ubishops.ca}
\affiliation{Physics Department and STAR Research Cluster, 
Bishop's University, 
Sherbrooke, Qu\'ebec, Canada J1M~1Z7
}
\author{Marianne Lapierre-L\'eonard}
\email{mlapierre12@ubishops.ca}
\affiliation{Physics Department, Bishop's University, 
Sherbrooke, Qu\'ebec, Canada J1M~1Z7}

\begin{abstract} 
We derive a formula describing the transformation of the 
Hawking-Hayward quasi-local energy under a conformal rescaling of 
the spacetime metric. A known formula for the transformation of the 
Misner-Sharp-Hernandez mass is recovered as a special case.
\end{abstract}


\maketitle

\section{Introduction}
\label{sec:1}

Historically, the 
notion of a physical 
energy contained in a compact 3 dimensional spacetime region (quasi-local energy) has proved to be rather complicated to 
identify and various 
attempts have been made to define a physically meaningful quantity
for this concept in general relativity (see \cite{Szabados} for a 
review). In the presence of spherical symmetry, however, (in 
general, without asymptotic flatness) the physical mass-energy is
commonly identified with the Misner-Sharp-Hernandez construct 
\cite{MSH}, the task becoming much more involved when the spacetime
is not spherically symmetric. There is a large consensus in the 
relativity community that, in the general case, the Hawking-Hayward 
quasi-local energy \cite{Hawking, Hayward} is an appropriate 
concept to define the mass contained in a compact region of spacetime.

Conformal mappings of spacetimes are widely used in cosmology and 
black hole physics ({\em e.g.}, \cite{various} and references 
therein), as well as in Penrose-Carter 
diagrams bringing 
spatial infinity to a finite distance in asymptotically flat 
spacetimes ({\em e.g.}, \cite{Wald}), not to mention their utility in alternative theories of gravity ({\em e.g.}, \cite{mybooks}).
The conformal transformation properties of other 
quasi-local energy concepts have been discussed in the literature:
see \cite{DeserTekin06} for the Arnowitt-Deser-Misner mass 
and \cite{BoseLohiya} for the Brown-York quasi-local energy. 
However, in spite of the importance of the Hawking-Hayward 
quasi-local energy, a transformation formula for this quantity 
under conformal spacetime mappings is not available in the 
literature except for the special case of spherical symmetry, in 
which the Hawking-Hayward mass reduces to the 
Misner-Sharp-Hernandez mass \cite{Hayward, Hayward2}, for which the 
transformation  was derived recently 
\cite{FaraoniVitagliano14}. The present paper fills this gap in the 
literature. It is necessary to first analyze how the various 
quantities appearing in the definition of the Hawking-Hayward mass
transform under conformal mappings, which is done in the next 
section. The following section derives the desired transformation 
property (eq.~(\ref{conh}), which is the main result of this 
paper). We then check that, in the special case 
of spherical symmetry, this formula reproduces the known one for 
the transformation of the Misner-Sharp-Hernandez mass.

\section{Hawking-Hayward quasi-local energy}
\label{sec:2}

The Hawking-Hayward mass $M_H$ \cite{Hawking, Hayward} is a 
functional of the 
spacetime metric $g_{ab}$ and of an  
embedded (spacelike, compact, and orientable) 2-surface $S$ 
defined by
\be 
M_H:=\frac{1}{8\pi}\sqrt{\frac{A}{16\pi}}\int_S 
\mu\left(\mathcal{R}+\theta_+\theta_- 
-\frac{1}{2}\sigma_{ab}^+\sigma^{ab}_-  
-2\omega_a\omega^a\right)\label{E:HH} 
\ee
where $\mathcal{R}$ is the induced Ricci scalar on $S$, 
$\theta_{\pm}$ 
and $\sigma_{ab}^{\pm}$ are the expansion and shear tensors of a 
pair of null  
geodesic congruences (outgoing and ingoing from the surface $S$), 
$\omega^a$ is the projection onto $S$ of the commutator of the null 
normal vectors to $S$,  $\mu$ is the volume 2-form on the surface 
$S$, and $A$ is the area of $S$.

\subsection{The surface $S$}

A co-dimension 2 surface $S$ in a 4-dimensional spacetime is 
defined as the set of points where 
two independent functions take constant values. Let us call these 
functions $\phi^{(1)}$ and $\phi^{(2)}$. Then,
\be
S:=\{x \,|\, \phi^{(1)}(x)=\phi^{(1)}_0\}\cap  \{x \,|\, 
\phi^{(2)}(x)=\phi^{(2)}_0\} \,.
\ee
Different 2-dimensional surfaces can be defined by choosing 
different constants $\phi^{(i)}_0$, indeed entire 2 parameter 
foliations of the 4-dimensional manifold can be defined in this way for 
certain choices of the functions $\phi^{(i)}$ in special 
circumstances.


Even before invoking a metric, we can associate two 1-forms with 
the surface $S$ by
\begin{align}
l_a&:=\partial_a \phi^{(1)} \,, \label{E:ln0}\\
n_a&:=\partial_a \phi^{(2)} \,, \label{E:ln}
\end{align}
whose restriction to the set $S$ defines the `normal direction' 1-forms -- since there is no 
metric yet we don't have a notion of orthogonality and we do 
not have a canonical way to define `normal vectors' by raising the 
indices on $l_a$ and $n_a$.   These 1-forms clearly satisfy 
\be
\nabla_{[a}l_{b]}=0=\nabla_{[a}n_{b]}
\ee
which are the conditions for the 1-forms $l$ and $n$ to be 
closed, $dl=0=dn$. This implies the metric-independent 
condition of hypersurface orthogonality on both $l$ and $n$
\be
n\wedge dn=0=l\wedge dl.
\ee

The condition of hypersurface orthogonality for a 1-form is more 
general than the closure condition: 1-forms proportional 
to the differential of a function, $fdg$ (where $g$ and $f$ are functions), are in general not closed but are hypersurface orthogonal
\be
\left(fdg\right)\wedge d\left(fdg\right)=fdg\wedge df\wedge dg=0\,.
\ee
That is, we are free to scale the 1-forms $l$ and $n$ 
by arbitrary functions without spoiling their hypersurface 
orthogonality.  

We note that there is also considerable freedom in the functions 
$\phi^{(i)}$ defining $S$ since any other set of two 
(sufficiently nice) independent functions 
$\psi^{(i)}(\phi^{(1)},\phi^{(2)})$ will 
define the same set $S$. 

Introducing now a metric $g_{ab}$, we impose that both $l$ and $n$ 
be 
null $g^{ab}l_a l_b=0=g^{ab}n_a n_b$ and from them form the scaled 1-forms 
\be
\overline{l}_a:=\frac{l_a}{\sqrt{-l\cdot n}}, \quad 
\overline{n}_a:=\frac{n_a}{\sqrt{-l\cdot n}} \,,
\ee
where $ l\cdot n \equiv g^{ab} l_a n_b$. In general, null normals 
do not have a canonical scaling in 
contrast to the timelike and spacelike cases where we can 
normalise to $\mp 1$. Our particular choice of scaling for the null 
normals implies that
\be
\overline{l}^a \overline{n}_a=-1
\ee
so that the induced metric on the surface $S$ is written as
\be
h_{ab}=g_{ab}+\overline{l}_a \overline{n}_b+\overline{l}_b 
\overline{n}_a.
\ee
We note also that in the null case there is a restricted freedom 
on the functions $\phi^{(i)}$; only un-mixed reparametrizations $\psi^{(i)}(\phi^{(i)})$ leave the null condition on the normals intact. 

One often finds in the literature the double-null construction 
based on the vectors $L^a$ and $N^a$ defined by
\be
L^a:=\frac{l^a}{l\cdot n}, \quad N^a:=\frac{n^a}{l\cdot n}
\ee
which have the property of Lie-dragging the 3-surfaces of constant 
$\phi^{(i)}$: 
\be
L(\phi^{(2)}) =1 = N(\phi^{(1)}) \,.
\ee
Despite appearances, this property is not sufficient to define the 
basis vectors $\partial/\partial \phi^{(i)}$ and would only do so 
in the case that the commutator of $L$ and $N$ is orthogonal to the 
surface $S$, that is $h_{ab}[ L ,N]^a= 0$,  a condition known as  
`surface-forming'. In general, the partial derivative 
$\partial/\partial \phi^{(i)}$ contains also a component tangent 
to $S$:
\be
\partial/\partial \phi^{(2)}= L+r  \quad 
\text{where} \quad h_{ab}r^a=0
\ee
and similarly for $N$. Such considerations play a role in the 
2+2 formulation of general relativity (see {\em e.g.} \cite{gour}).

Note that while $l$ (and $n$) is tangent to an affinely 
parametrized geodesic, 
\begin{equation}
l^a\nabla_a l_b=l^a \nabla_b l_a 
= \frac{1}{2} \nabla_b\left(l_a l^a\right)=0 \,,
\end{equation}
$\overline{l}^a$ (nor $\overline{n}^a$) is not, but is still 
tangent to a non-affinely parametrized geodesic. Defining 
$\text{e}^{m}:=\sqrt{-l\cdot n}$ we have 
\begin{align}
\overline{l}^a \nabla_a \overline{l}^b&=\text{e}^{-m} l^a\nabla_a 
\left( \text{e}^{-m} l^b \right) \notag\\
&=\text{e}^{-m}l^a l^b\nabla_a \text{e}^{-m}+\text{e}^{-2m}l^a \nabla_a l^b \notag\\
&=-\text{e}^{-2m}l^a l^b\nabla_a m\notag\\
&=\left(-\overline{l}^a\nabla_a m\right)\overline{l}^b \notag\\
&= \left( -\mathcal{L}_{\overline{l}} m \right)  \overline{l}^b
\end{align}
where $\mathcal{L}$ is the Lie derivative. A similar result holds 
for $\overline{n}^a$. 

The quantities appearing in the definition of the Hawking-Hayward 
energy are defined in terms of Lie derivatives in the directions 
$\overline{l}$ and $\overline{n}$. The expansions $\theta_\pm$ and  
shears $\sigma_{ab}^\pm$ are defined by
\begin{align}
\theta_{+}&:=\frac{1}{2} \, h^{ab} \mathcal{L}_{\overline{l}}  
\,\,h_{ab} \,, \\
\theta_{-}&:=\frac{1}{2} \, h^{ab} \mathcal{L}_{\overline{n}}  
\,\,h_{ab} \,, \\
\sigma^{+}_{ab}&:=h^c_a h^d_b \mathcal{L}_{\overline{l}}\,\, 
h_{cd}-h_{ab}\theta_+  \,,\\
\sigma^{-}_{ab}&:=h^c_a h^d_b \mathcal{L}_{\overline{n}}\,\, 
h_{cd}-h_{ab}\theta_-  \,,
\end{align}
while the anholonomicity $\omega_a$ is defined by the commutator 
of $l^a$ and $n^a$ (or $\overline{l}^a$ and $\overline{n}^a$):
\be
\omega_a:=  \frac{1}{2}\, \frac{1}{l\cdot n}h_{ab} 
\left[l,n\,\right]^{b}= \frac{1}{2}\,\frac{1}{\overline{l}\cdot\overline{n} } h_{ab} 
\left[\overline{l}, \overline{n}\,\right]^{b} \,,
\ee
where $[\, , \,]$ is the Lie bracket. Note that the anholonomicity is invariant under rescaling of the normals. These definitions are 
textbook-standard except for the anholonomicity $\omega_a$ which deserves 
special mention. 
Usually\footnote{For example see Refs.~\cite{carroll} or 
\cite{poisson}.}  
when one is dealing with a single null congruence $v^a$, one 
introduces an auxiliary null vector $u^a$ such that $v^a u_a=-1$ 
and defines a 2-form $\omega_{ab}$ called the twist form by 
$h^d_{[a} h^c_{b]}\nabla_c v_d$. This object is important because 
it vanishes if and only if $v^a$ is hypersurface-orthogonal (it is 
proportional 
to the gradient of a scalar).  This object is not independent of 
the auxiliary vector \cite{carroll} but its square 
$\omega_{ab}\,\omega^{ab}$ is. One can show that if 
$\omega_{ab}=0$ for some choice of $u^a$, then it is zero for all 
choices of $u^a$, so the vanishing of $\omega_{ab}$ is a 
$u^a$-independent condition \cite{poisson} and hence hypersurface 
orthogonality does not depend on the choice of auxiliary $u^a$.  
What we have here is somewhat different. We are dealing here with 
a single 2-surface which is the intersection of two null 
3-surfaces. The vectors $l^a$ and $n^a$ certainly are 
hypersurface-orthogonal individually to these null 3-surfaces and 
hence both possess vanishing $\omega_{ab}$, but this is not what 
the 
anholonomicity $\omega_a$ is measuring (otherwise it would 
trivially vanish in this 2+2 formulation). The anholonomicity 
describes how the two $l^a$ and $n^a$ vectors `weave together' 
(integrate to) a 2-dimensional hypersurface, whether they are really tangent 
to a genuine 2D submanifold. Now this `integral' sub-manifold, if 
it exists, is not $S$, but the orthogonal complement $ S^\perp $ 
since the $l^a$ and $n^a$ vectors are orthogonal to $S$.


\section{Transformation of the Hawking-Hayward energy under 
conformal mappings}
\label{sec:3}

We now consider the relationship between $M_H$ defined with respect to two different metrics $g_{ab}$ and $\tilde{g}_{ab}$ which are related by a conformal factor
\be
\tilde{g}_{ab}=\Omega^2 g_{ab}.
\ee
Of course, since the surface $S$ is defined independently of the metric, we are considering how the energy content contained within a single surface changes when one scales the metric as above.

In the expression~(\ref{E:HH}) (with respect to the metric $\tilde{g}$) of 
the Hawking-Hayward mass of a 2-surface $S$ in the  
conformally scaled world, the area of this surface is
\be
\tilde{A}=\int_{S} \tilde{\mu} 
=\int_{S} d^2x \sqrt{\tilde{h}} 
=\int_{S} d^2x \sqrt{\Omega^4 h} 
=\int_{S} \mu \, \Omega^2  \,.
\ee
Unless $\Omega $ is constant on $S$, one 
cannot extract it from the integral and we have
\be
\sqrt{ \tilde{A}} =\sqrt{ \frac{ \int_{S} 
\mu \, \Omega^2 }{A}} \, \sqrt{A} \,.
\ee
Only when $\Omega$ is constant on ${S}$ can we write $\sqrt{ \tilde{A}}=\Omega \sqrt{A}$. This expression will 
be used in the integral defining the quasi-local energy.

We will make use in what follows of the contracted Gauss equation \cite{Hayward}
\be
{\cal R}+\theta_{+} \theta_{-} -\frac{1}{2} \, \sigma_{ab}^{+} 
\, \sigma^{ab}_{-} =h^{ac} h^{bd} R_{abcd}
\ee
which provides a very useful expression of a part of the Hawking-Hayward energy in terms of simpler quantities. 
In the conformally rescaled world we have
\be
\tilde{{\cal R}}+ \tilde{\theta}_{+} \tilde{\theta}_{-} 
-\frac{1}{2} \, \tilde{\sigma}_{ab}^{+} \,  
\tilde{\sigma}^{ab}_{-} = \tilde{h}^{ac} 
\tilde{h}^{bd} \tilde{R}_{abcd} \,.
\ee
\begin{widetext}
Using the fact that $\tilde{h}^{ab}=\Omega^{-2} h^{ab}$ and 
the well known transformation property of the Riemann 
tensor under conformal transformations ({\em e.g.}, \cite{Wald}, 
p.~466) 
\begin{align}
 \widetilde{ {R_{abc}}^d}  & =  {R_{abc}}^d  
+ 2 \delta_{[a}^d \nabla_{b]}\nabla_c \ln \Omega 
- 2 g^{de} g_{c[a}\nabla_{b]}\nabla_e \ln \Omega 
 + 2 \nabla_{[a}\ln \Omega \, \delta_{b]}^d\nabla_c 
\ln \Omega \notag\\
&\hspace{5cm}- 2 \nabla_{[a}\ln \Omega g_{b]c} g^{de}\nabla_e \ln \Omega - 2 g_{c[a}  \delta_{b]}^d g^{ef} \nabla_e \ln \Omega 
\, \nabla_f \ln \Omega 
\end{align}
and, lowering one index, 
\begin{equation}
\tilde{R}_{abcd}  =  \tilde{g}_{ds} \tilde{R}_{abc}{}^s = \Omega^2 g_{ds}\tilde{R}_{abc}{}^s
\end{equation}
it follows that
\begin{align}
 \tilde{h}^{ac} \tilde{h}^{bd} \tilde{R}_{abcd} &=
\Omega^{-2} \left( {\cal R}+\theta_{+} \theta_{-} -\frac{1}{2} 
\, \sigma_{ab}^{+} \, \sigma^{ab}_{-} \right)
+2\Omega^{-2} \left[ 
h^{ac} h^b_s \delta^s_{[a} \nabla_{b]} \nabla_c\ln \Omega 
- h^{ac} h^{be} g_{c[a} \nabla_{b]} \nabla_{c}\ln \Omega
\right.
 \notag \\
&\hspace{15mm}\left. +h^{ac} h^b_s \nabla_{[a} \ln \Omega \, \delta^s_{b]} 
\nabla_{c}\ln \Omega
- h^{ac} h^{bd} \nabla_{[a} \ln \Omega \, g_{b]c} 
\nabla_d\ln 
\Omega 
 - h^{ac} h^b_s g_{c[a}\delta^s_{b]} 
\nabla^f \ln \Omega \, \nabla_f \ln \Omega \right] \,,
\label{Delta}
\end{align}
where $h^a_b$ is a 2-dimensional Kronecker delta. 
By computing the terms appearing on the right hand side of 
this equation, one obtains
\begin{align}
&h^{ac} h^b_s \delta^s_{[a} \nabla_{b]} \nabla_c\ln \Omega 
-h^{ac} h^{be} g_{c[a} \nabla_{b]} \ln \Omega = - h^{ab} \nabla_a \nabla_b \ln \Omega \,,\\
&h^{ac} h^b_s \nabla_{[a} \ln \Omega \delta^s_{b]}
\nabla_{c}\ln \Omega 
-h^{ac} h^{bd} \nabla_{[a} \ln \Omega \, g_{b]c}
\nabla_{d}\ln \Omega 
= h^{ab} \nabla_a \ln \Omega \nabla_b \ln \Omega \,,\\
&h^{ac} h^b_s \, g_{c[a}\delta^s_{b]}
\nabla^f \ln \Omega \, \nabla_f \ln \Omega =
\frac{ \nabla^c \Omega \nabla_c \Omega}{\Omega^2} \,.
\end{align}
Putting everything together gives 
\begin{align}
\tilde{ {\cal R}} + \tilde{\theta}_{+} \tilde{\theta}_{-} 
-\frac{1}{2}
\, \tilde{\sigma}_{ab}^{+} \, \tilde{\sigma}^{ab}_{-} 
&= \Omega^{-2} 
\left( {\cal R}+\theta_{+} \theta_{-} -\frac{1}{2} \, 
\sigma_{ab}^{+} \, \sigma^{ab}_{-} \right) \notag \\
& +2\Omega^{-2}  h^{ab} \left( \nabla_a \ln \Omega  
\, \nabla_b \ln \Omega - \nabla_a \nabla_b \ln \Omega 
\right) 
-2\Omega^{-2} \nabla^c \ln \Omega \, \nabla_c \ln \Omega 
\,.
\end{align}
Using now the fact (which we will prove below) that 
\be
\tilde{\omega}_a \tilde{\omega}^a =\Omega^{-2} 
\omega_a \omega^a \,,
\ee
the integrand in the Hawking-Hayward quasi-local energy is 
seen to transform as 
\begin{align}
  \tilde{ {\cal R}}+\tilde{ \theta}_{+} \tilde{\theta}_{-} 
-\frac{1}{2}\, \tilde{\sigma}_{ab}^{+} \, 
\tilde{\sigma}^{ab}_{-}
-2 \, \tilde{\omega}_a \tilde{\omega}^a
&= \Omega^{-2} 
\left( {\cal R}+\theta_{+} \theta_{-} -\frac{1}{2} \, 
\sigma_{ab}^{+} \, \sigma^{ab}_{-} 
-2 \, \omega_a \omega^a \right)\notag \\
&+2\Omega^{-2} \left[  h^{ab} \left( \frac{ 
2 \nabla_a \Omega \nabla_b  \Omega}{\Omega^2}  
- \frac{\nabla_a \nabla_b \Omega}{\Omega} \right) \right. 
 \left. - \frac{ \nabla^c  \Omega \nabla_c 
\Omega}{\Omega^2} \right] \,.
\end{align}
The Hawking-Hayward quasi-local energy itself then 
transforms according to 
\begin{align}
\tilde{M}_H &=\frac{1}{8\pi} \sqrt{ \frac{\tilde{A}}{16\pi}} 
\int_{S} \mu \left( {\cal R} +\theta_{+} \theta_{-} 
-\frac{1}{2} \, \sigma_{ab}^{+} \sigma^{ab}_{-}
-2 \, \omega_a {\omega}^a \right)+ \frac{1}{4\pi} \sqrt{ \frac{\tilde{A}}{16\pi}}
\int_{S} \mu \left[ h^{ab} 
\left( \frac{
2 \nabla_a \Omega \nabla_b  \Omega}{\Omega^2}
- \frac{\nabla_a \nabla_b \Omega}{\Omega} \right) - \frac{ \nabla^c  \Omega \nabla_c
\Omega}{\Omega^2} \right] \notag \\
&=\sqrt{\frac{\tilde{A}}{A}}M_H+ \frac{1}{4\pi} \sqrt{ \frac{\tilde{A}}{16\pi}}
\int_{S} \mu \left[ h^{ab} 
\left( \frac{
2 \nabla_a \Omega \nabla_b  \Omega}{\Omega^2}
- \frac{\nabla_a \nabla_b \Omega}{\Omega} \right) - \frac{ \nabla^c  \Omega \nabla_c
\Omega}{\Omega^2} \right]
\label{conh}.
\end{align}
\end{widetext}

One can also arrive at this final result \eqref{conh} by 
transforming directly the ingredients from which $M_H$ is 
constructed. The 1-forms $l_a$ and $n_a$ in 
eqs.~(\ref{E:ln0}) and~(\ref{E:ln}) are 
defined independently of the metric and therefore are unchanged 
under conformal transformations. 
\be
\tilde{l}_a=l_a, \quad \tilde{n}_a=n_a.
\ee
Hence
\begin{align}
\tilde{\overline{l}}_a &:=\frac{\tilde{l}_a}{\sqrt{ 
-\tilde{g}^{bc}\tilde{l}_b  
\tilde{n}_c}} \notag\\
&=\frac{l_a}{\sqrt{-\Omega^{-2}g^{bc}l_b  
n_c }}\notag\\
&=\Omega \, \frac{l_a }{\sqrt{-g^{bc }l_b  
n_c }}\notag\\
&:=\Omega \,\overline{l}_a 
\end{align}
and, similarly, $ \tilde{\overline{n}}_a =\Omega \,\overline{n}_a$. 
We also have
\be
\tilde{\overline{l}}^a =\Omega^{-1} \,\overline{l}^a \,, \quad 
\tilde{\overline{n}}^a =\Omega^{-1} \, \overline{n}^a \,.
\ee
Note that in the conformal world we still have the condition
\be
\tilde{\overline{l}}^c \tilde{\overline{n}}_c =-1
\ee
so that the metric is still expressed as
\be
\tilde{h}_{ab }=\tilde{g}_{ab} 
+\tilde{\overline{l}}_a \tilde{\overline{n}}_b  
+\tilde{\overline{l}}_b \tilde{\overline{n}}_a \,.
\ee

As an aside, and for ease of computation, it is easy to show that
\begin{align}
\theta_+\theta_- 
-\frac{1}{2}\sigma_{ab}^+\sigma^{ab}_-&=\notag\\
\hspace{-3mm}&\frac{1}{2} \left[h^{ab} h^{AB}-h^{a A}h^{b B}\right] 
\left(\mathcal{L}_{\overline{l}}h_{ab} 
\right)\left(\mathcal{L}_{\overline{n}}h_{AB}\right)
\end{align}
whence it follows that
\begin{align}
\tilde{\theta}_+\tilde{\theta}_- 
-\frac{1}{2}\tilde{\sigma}_{ab}^+ 
\tilde{\sigma}^{ab}_- 
& =\Omega^{-2}\left(\theta_+\theta_--\frac{1}{2} 
\sigma_{ab}^+\sigma^{ab}_-\right)\notag\\
&+\Omega^{-6}\left(\overline{l}^a \nabla_a \Omega^2\right) 
\left(\overline{n}^b \nabla_b \Omega^2\right)\notag \\
&+\frac{2}{\Omega^{3}}\theta_{\overline{l}}\left(\overline{n}^a 
\nabla_a \Omega\right)+\frac{2}{\Omega^{3}}\theta_{\overline{n}} 
\left(\overline{l}^b \nabla_b \Omega\right) \, . \label{E:cross}
\end{align}

The Ricci scalar of the 2-surface $S$ transforms under the 
conformal transformation as
\be
\tilde{\mathcal{R}}=\Omega^{-2}\mathcal{R} 
-\frac{2}{\Omega^3}h^{ab} 
D_a D_b \Omega + \frac{2}{\Omega^4}h^{ab} 
\left(D_a \Omega\right)\left(D_b \Omega\right) \,,
\ee
where $D_a$ is the covariant derivative on $S$ which is defined 
with respect to the metric $h_{ab}$ and whose action is related to the 
4-dimensional  $\nabla_a$ by, using the example of a (1,1)-tensor, 
\be
D_c X^a_b :=h^d_c h^a_f  h_b^e 
\nabla_d X^f_e \,. 
\ee
We can re-express everything in terms of the 4D covariant 
derivative as follows
\begin{align}
D_a \Omega&=h^c _a \nabla_c \Omega  \,, \\
D_a D_b \Omega &=h^c_a h^d_b \nabla_c D_d 
\Omega \,, \notag\\
&=h^c_a h^d_b \nabla_c \left( h^f_d\nabla_f \Omega \right)
\end{align}
so that
\begin{eqnarray}
h^{ab}D_a D_b \Omega &=& h^{ab} 
h^c_a h^d_b\nabla_c \left( h^e_d\nabla_e 
\Omega \right) \notag\\
&&\nonumber\\
&=& h^{cd}\nabla_c\left[ \delta^e_d 
+\left(\overline{l}^e \overline{n}_d +\overline{n}^e \overline{l}_d \right) 
\right]\nabla_e \Omega\notag\\
&&\nonumber\\
&=& h^{cd}\nabla_c \nabla_d
\Omega+\left(\overline{l}^e\nabla_e 
\Omega\right)h^{cd}\nabla_c \overline{n}_d \notag\\
&&\nonumber\\
&\,& + \left(\overline{n}^e\nabla_e 
\Omega\right)h^{cd}\nabla_c \overline{l}_d \notag\\
&&\nonumber\\
&=& h^{cd}\nabla_c \nabla_d 
\Omega+\left(\overline{l}^e\nabla_e 
\Omega\right)\theta_{\overline{n}}+ 
\left(\overline{n}^e \nabla_e  
\Omega\right)\theta_{\overline{l}}\,. \nonumber\\
&&
\end{eqnarray}
In the last line we have made use of the equivalent definition for the expansion $\theta_{\overline{n}}:=h^{ab}\nabla_a \overline{n}_b$ and similarly for $\theta_{\overline{l}}$.  We are then left with
\begin{eqnarray}
\tilde{\mathcal{R}} 
&=& \frac{\mathcal{R}}{\Omega^{2}} -\frac{2}{\Omega^3}h^{ab} 
\nabla_a \nabla_b \Omega + \frac{2}{\Omega^4}h^{ab} 
\left(\nabla_a \Omega\right)\left(\nabla_b \Omega\right)\notag\\
&&\nonumber\\
&\,& - 
\frac{2}{\Omega^3}\left(\overline{l}^e\nabla_e  
\Omega\right)\theta_{\overline{n}}-\frac{2}{\Omega^3}   
\left(\overline{n}^e\nabla_e  
\Omega\right)\theta_{\overline{l}} \,.\label{xxx43}
\end{eqnarray}
Note the cancellation of the cross terms in $\theta_{\bar{l}}$ and 
$\theta_{\bar{n}}$ when combining eqs.~\eqref{xxx43} and \eqref{E:cross}.

We also have
\begin{align}
\tilde{\omega}_a &:=  
\frac{1}{2 \tilde{l}\cdot\tilde{n}}\tilde{h}_{ab} 
\left[\tilde{l}, \tilde{n} \right]^a \notag\\
&= \frac{1}{2\,\Omega^{-2}l\cdot n} 
\Omega^{2}h_{ab}\left[ 
\Omega^{-2}\overline{l}, \Omega^{-2}\overline{n}\right]^a\notag\\ 
&= \frac{1}{2\,l\cdot n} 
h_{ab}\left[ l, n\right]^a =\omega_a 
\end{align}
where we have used $\tilde{l}^a=\tilde{g}^{ab}\tilde{l}_b=\Omega^{-2} g^{ab} l_b$.  Combining these terms we arrive at the result \eqref{conh} obtained above.

\section{Spherical symmetry}
\label{sec:4}

As a check of the formulae derived above, it is useful to 
discuss spherically symmetric situations. In spherical 
symmetry the Hawking-Hayward quasi-local energy reduces 
\cite{Hayward, Hayward2} to 
the Misner-Sharp-Hernandez mass \cite{MSH}. The 
transformation property of the Misner-Sharp-Hernandez mass 
under conformal mappings was 
recently reported in Ref.~\cite{FaraoniVitagliano14}. A 
spherically symmetric metric can be put in the form
\be\label{sphericalmetric}
ds^2 =-A(t,R) dt^2 +B(t,R) dR^2 +R^2 d\Omega_{(2)}^2 \,,
\ee
where $d\Omega_{(2)}^2=d \theta^2 +\sin^2 \theta 
\, d\varphi^2$ 
is the line element on the unit 2-sphere and $R$ is the 
areal radius. Under a conformal transformation with 
conformal factor $\Omega=\Omega (t, R)$ (to preserve 
spherical symmetry), this line element becomes $
d\tilde{s}^2 =\Omega^2 ds^2 $ and the areal radius is 
$\tilde{R}=\Omega\, R$. The Misner-Sharp-Hernandez mass in 
the rescaled world is simply \cite{FaraoniVitagliano14}
\be \label{MSHtransformation}
\tilde{M}_H=\Omega M_H -\frac{R^3}{2\Omega} \, \nabla^c\Omega 
\nabla_c\Omega -R^2 \nabla^c\Omega \nabla_c R \,.
\ee
In spherical symmetry, a sphere $S$ of  constant radius $R$ 
has area $A=4\pi R^2$, $\sqrt{ \tilde{A}} =\Omega 
\sqrt{A}$, and $h^{ab}=\, \mbox{diag} \left( R^{-2}, R^{-2} 
\sin^{-2} \theta \right)$. In order to check whether 
eq.~(\ref{conh}) reproduces eq.~(\ref{MSHtransformation}) 
correctly, one needs to compute the various terms in the 
integrand of eq.~(\ref{conh}). The integrand is constant 
over a 
sphere of constant radius and the integral reduces to the 
product of the integrand and the area $4\pi R^2 $ of this 
sphere. We have
\begin{align}
\frac{\Omega}{8\pi} \sqrt{ \frac{A}{16\pi}} 
\int_{\tilde{S}} \mu
&\left( {\cal R}+\theta_{+} \theta_{-} -\frac{1}{2} \, 
\sigma_{ab}^{+} \sigma^{ab}_{-} -2\omega_a \omega^a 
\right) \notag \\
&\hspace{30mm}=\Omega M_H \,,
\end{align}
and
\be
\frac{\Omega}{8\pi} \sqrt{ \frac{A}{16\pi}} 
\int_{\tilde{S}} \mu  \left( 
-\frac{2\nabla^c\Omega\nabla_c\Omega}{\Omega^2} \right)=
-\frac{R^3}{2\Omega} \, \nabla^c\Omega \nabla_c\Omega \,.
\ee
Since $\Omega=\Omega(t, R)$ we have $\nabla_a \Omega 
=\dot{\Omega} \, \delta_{a0}+\Omega' \, \delta_{a1}$, where 
a 
dot and a prime denote differentiation with respect to $t$ 
and $R$, respectively. Then it is $
h^{ab}\nabla_a\Omega \nabla_b\Omega = 0 $ 
and we are left only with the quantity $ -h^{ab} \nabla_a 
\nabla_b \Omega $ to compute. We have 
\begin{eqnarray}
h^{ab} & \nabla_a &  \nabla_b \Omega =
h^{ab} \partial_a \partial_b \Omega -h^{ab} \Gamma_{ab}^c 
\partial_c \Omega  \nonumber\\
&&\nonumber\\
&&\hspace{-0.7cm}= h^{22} \partial_{\theta\theta}\Omega +
h^{33} \partial_{\varphi\varphi}\Omega
- \left( h^{22} \Gamma^c_{22} + h^{33} \Gamma^c_{33} 
\right) \partial_c \Omega  \nonumber\\
&&\nonumber\\
&&\hspace{-0.7cm}=
-\frac{1}{R^2} \left( \Gamma^0_{22} \dot{\Omega} +
\Gamma^1_{22} \Omega' \right) 
-\frac{1}{R^2\sin^2 \theta} \left( \Gamma^0_{33} 
\dot{\Omega} + \Gamma^1_{22} \Omega' \right).\nonumber\\
&&
\end{eqnarray}
Using the line element~(\ref{sphericalmetric}), the 
Christoffel symbols are easily computed:
\begin{eqnarray}
\Gamma^0_{22} &=& \Gamma^0_{33}=  0 \,,\\
 &&\nonumber\\
\Gamma^1_{22} &=& -\frac{R}{B} \,, \;\;\;\;\;
\Gamma^1_{33} = -\frac{R\sin^2\theta }{B} \,,
\end{eqnarray}
which yields
\be
h^{ab} \nabla_a \nabla_b \Omega =\frac{2\Omega'}{BR} 
\ee
and
\begin{align}
\frac{R}{8\pi} \int_{\tilde{S}}\mu h^{ab} &\left( \frac{ 
\nabla_a\Omega \nabla_b \Omega}{\Omega} - \nabla_a 
\nabla_b\Omega \right) \notag\\
&=  
-\frac{R}{8\pi} (4\pi R^2) h^{ab} \nabla_a \nabla_b\Omega \notag\\
&= -\frac{R^2 \Omega'}{B} \,.
\end{align} 
This is the third term on the right hand side of 
eq.~(\ref{MSHtransformation}); in fact, 
\be
-R^2 \nabla^c\Omega \nabla_cR= -R^2 g^{ab} \nabla_a \Omega 
\, \delta_{b1}= -\frac{R^2}{B}\, \Omega' 
\ee
and the transformation property~(\ref{MSHtransformation}) 
of the Misner-Sharp-Hernandez mass is indeed  
a special case of eq.~(\ref{conh}).

\section{Discussion}
\label{sec:5}

Contrary to other notions of quasi-local energy, the transformation 
property of the Hawking-Hayward quasi-local energy is not discussed 
in the literature. This concept requires one to analyze the two 
null (outgoing and ingoing) geodesic congruences associated with a  
2-surface $S$, and to compute their optical scalars and 
anholonomicity. We have presented the relevant calculations, taking 
care to impose the normalization of the null normals used 
in the Hawking-Hayward quasi-local energy definition \cite{Hawking, Hayward}, in 
both the original and the conformally rescaled worlds. Putting 
together the various pieces, one arrives at the desired 
transformation property~(\ref{conh}) for $M_H$. The result is reproduced by the calculation which involves the contracted Gauss equation. In the special case of spherical symmetry, in which the 
Hawking-Hayward quasi-local energy reduces to the 
Misner-Sharp-Hernandez mass \cite{Hayward, Hayward2}, we recover  
a previous result on the transformation of this quantity under 
conformal rescalings \cite{FaraoniVitagliano14} further confirming our general result. 

Given the wide use of conformal transformations in cosmology and 
black hole physics, we expect the result obtained here to be useful  
in these areas. An application to cosmological perturbations will 
be reported elsewhere \cite{VMA}.

\begin{acknowledgments} 
This research is supported by Bishop's University and by 
the Natural Sciences and Engineering Research Council of 
Canada ({\em NSERC}). VV is supported by FCT-Portugal through
the grant SFRH/BPD/77678/2011.
\end{acknowledgments}



\end{document}